# Dynamic data models: an application of MOP-based persistence in Common Lisp

Pierre THIERRY

Independent developer
Thierry Technologies

pierre@thierry-technologies.com

Simon E.B. THIERRY
Ph.D. candidate
LSIIT, UMR CNRS-ULP 7005
Pôle API, BP 10413
Illkirch Cedex France
thierry@lsiit.u-strasbg.fr

**Abstract**

The data model of an application, the nature and format of data stored across executions, is typically a very rigid part of its early specification, even when prototyping, and changing it after code that relies on it was written can prove quite expensive and error-prone.

Code and data in a running Lisp image can be dynamically modified. A MOP-based persistence library can bring this dynamicity to the data model. This enables to extend the easy prototyping way of development to the storage of data and helps avoiding interruptions of service. This article presents the conditions to do this portably and transparently.

## 1 Introduction

Many applications provide a continuous service but cannot guarantee that they will run continuously. These applications must serialize and write part of their state to be able to restore it when they are executed again. Some of these applications are designed to run continuously but one cannot avoid to shut them down occasionally, *e.g.* because of hardware failure, maintenance or software updates, like network servers. Others are only needed part of the time and are designed to be executed only then, like accounting software. Occasionally an application is designed to be executed in a short time and only handle user input and hardcoded data but, as the use conditions evolve, it becomes necessary to stop its execution and be able to resume it later.

The developer must guarantee that a saved state will be readable and that the application will be able to effectively resume its execution, which creates a rather inconvenient set of requirements:

1. the format used must be able to represent arbitrary data

2. the format used must have an unambiguous syntax

3. the format used should be parsed efficiently

4. stored data must be kept synchronized with the code handling the state

5. the nature and structure of data must be able to evolve

Note that the fifth requirement actually implies a sixth corollary requirement: any data stored before a change in data structures has to be migrated from the older to the newer structures.

We can observe that the first three requirements are rather easily solved by well-known solutions produced by the fields of parsing and databases management. Yet, since it includes the fact the organization of data and code may change, the complete set of requirements constitutes a much harder problem [5].

This ability of an application to exhibit the behaviour of a continuous execution even when its execution is actually discontinuous is what one calls persistence. We will show how a library using the MetaObject Protocol (MOP) of Common Lisp can provide persistence at the cost of practically no effort from the developer, and fulfill the mentioned requirements. We will also show under which conditions such a transparent addition to the application can be made portably with the MOP.

The rest of this paper is organized as follows: section 2 shows the benefits of a dynamic data model; section 3 describes other approaches to make an application persistent; section 4 details the requirements and issues of a portable MOP-based transparent persistence library; section 5 shows how such a library integrates the data model in classical uses of Lisp dynamicity; section 6 sketches our use of a persistent library and how the library could be enhanced to better fit our requirements.

## 2 Data model flexibility spectrum

We will now inspect the design space of the solutions fulfilling those requirements according to their flexibility, from the most rigid to the most flexible.

The easiest solution is to build the data model and set it in stone before any code handling the data is written. This solution is widely used because it seems to dodge the issues of synchronization and migration of data. But when this fails because the data model has to be changed nonetheless, this solution backfires severely: the data model presumed immutability encourages tight coupling between the code and data structures holding serialized data as well as fine tuning of the format, both at least for optimization; thus any change typically proves to be not only very costly, as many code portions have to be changed, but also very error-prone, as the lack of abstraction makes it easier to miss code that had to be updated.

Between this solution and the next one in the spectrum of flexibility, there is a wide range of solutions using a more or less well defined API to abstract the access to serialized data. Although they are obviously better than an *ad hoc* code, they actually only avoid the coupling between the code and the format used to serialize data. The developer still has to ensure that code and data are consistent and data is migrated to newer structures.

To avoid this burden, a solution is to define both in-memory and serialized data structures from the same source. This is typically achieved by using a Domain Specific Language (DSL) [12] for the purpose, from which code is compiled that deals with data structures to create them, access them, write them and read them back when needed. This is a widely used solution in the Java world, as seen in projects such as JBoss Hibernate or Oracle TopLink and as specified in Sun's JDO.

This doesn't solve the migration issue in itself, but makes it possible to do so: when data structures are defined this way, it becomes easy to develop an automatic tool which, given the older and the newer definitions, migrates the data. Still, this solution has all the classical problems of a DSL [11]: it is a non-trivial issue to get the DSL right, in term of expressiveness, and the more expressive it is, the more the developer has to learn to use it. It moves the burden from having to keep things up to date to learning a new language, which is a step forward, but not a complete solution. Moreover, depending on the DSL, the developer could want to move on to a data model that cannot be expressed with the DSL. This could force the developer to stop using the DSL, and rebuild everything from scratch. If he later discovers that with further modifications, he can use the DSL again, mental care may blow the budget.

Because of the remaining burden that learning and using a DSL for data model definition imposes, an interesting solution is the use of a persistence library. A persistence library basically provides code to serialize arbitrary data structures used in an application. But having to call such a library explicitly both cripples the code and is a source of hard to find bugs (because they can manifest themselves from obsolete data, typically causing subtly wrong results). The programming language implementation is able to call the persistence library safely, ensuring data is properly deserialized before read and serialized after write, and the easiest way to leverage this ability is the use of a transparent persistence library, which should provide access to persistent data structures with the same syntax used to access built-in data structures.

In the case of a portable transparent persistence library, if the semantics of persistent data structures are consistent with the semantics of built-in data structures, it may be possible to turn an application into a persistent one merely by activating persistence (that is, opening a store and initializing it if necessary) and declaring some objects persistent (either individually or for the extent of some class). This even makes it possible to use the same application in a transient or persistent way according to the conditions.

We will show that Common Lisp's MOP provides precisely the framework needed to portably implement a transparent persistent library for an object-oriented application.

## 3 Related works

Making an application persistent can be achieved from within the application, which is our approach, or by orthogonal persistence, that is persistence provided outside the application.

Orthogonal persistence has been a research area in OS design and has been implemented in systems such as KeyKOS [16], EROS [24] or Coyotos [22], where it simplifies security reasoning [23] as well as provides system-wide backup, fail-over support and extremely short restart times [7]. OS-wide orthogonal persistence also typically avoids the implementation difficulties of non-orthogonal persistence, as no serialization actually takes places: with the appropriate design, memory pages can be stored as-is and restored without modification. Thus it is typically both very flexible and efficient, albeit absolutely not portable, but tied to a specific system architecture.

When not provided by the OS, orthogonal persistence can be provided by the programming language implementation. The first language to do so was PS-Algol [3,10], where persistence was demonstrated to provide, along with closures, a sufficient basis to implement modules or separated compilation [1,2]. With the intent to exhibit Common Lisp's virtues with the additional benefits of persistence, a persistent implementation of Common Lisp, UCL+P [13,14], was designed. While providing a great flexibility, orthogonal persistence at the language level also ties a design to a specific system. Though the language implementation may be portable, the application in itself isn't.

In Common Lisp particularilly, persistence has been an active field with various libraries. As Common Lisp is favoured by its users for its flexibility, vendors provide transparent persistence for their implementation, like Statice for the Genera OS used on Lisp Machines or AllegroCache for Allegro CL.

The MOP makes it possible to provide transparent persistence with portable code. The PCLOS library was written with the intent to demonstrate this possibility [17–19], though this library doesn't enable class redifinition for persistent classes. Adding transparent persistence, although arguably difficult to implement comprehensively and efficiently, is even a classical example of MOP use [9]. Not all MOP-based persistence libraries handle the storage themselves, though: PLOB!, for instance, relies on an object-oriented database coded in C to store persistent objects.

While most persistence libraries offer the ability for the application to operate on a data set possibly larger than memory, some, for the sake of speed, take the approach of prevalence [6], where all the data set is present in memory, like CL-Prevalence or BKNR. While providing speed and simplicity, prevalence is of interest only in a limited subset of the persistence use cases.

## 4 Implementation issues

This section presents the requirements of a portable implementation of a MOP-based persistence library enabling our approach.

### 4.1 Language requirements

The implementation can be divided in two major parts, the serializer and the transparency mechanism.

The serializer is conceptually a bijective function that maps objects of the programming language to their representation as a byte string; the deserializer is its inverse. Its implementation includes the following components: a serializer for built-in types of the language and one for user-defined data structures.

The serialization of built-in types has very few requirements: only polymorphism is actually needed. In a statically typed language, early binding should be able to dispatch to the relevant serialization function. If the statically typed language allows type-unsafe data to be used – like references cast to void pointers in C or C++, serialization might not be possible in the general case. In a dynamically typed language, late binding or type information will be necessary.

Some types may be impossible to serialize at all, or at least portably. Without detailed metadata, references to services of the host OS will not be serial-

izable – *e.g.* a file handle without the file's path. Even with proper metadata, serialization of such types may be inaccurate: the services exposed through the reference might have their state modified when the application is not running and thus impossible to receive notification. Deserialization of these types might need to raise exceptions and try and simulate such notification.

Executable code is also a source of difficulty for serialization, as functions or procedures are typically opaque objects. Persistent closures, though, have been shown to be particularly interesting [1,2]. As knowing wether a procedure is serializable without any other ability than to execute it is undecidable, one needs a facility from the language: either metadata from which the procedure can be rebuilt – *e.g.* its source code – or inspection of its code.

Serialization of user-defined types is built on top of serialization of built-in types, with reflection. That is, the serializer must be able to inspect the components of any user-defined object. The serialization then consists in the identification of the user-defined type and the serialization of each component.

Thus, the basic requirements for serialization are some kind of polymorphism, depending on the type system of the language, and reflection for compound data. Some built-in types may need specific facilities to be serializable.

Another concern is object identity preservation. As classical solutions have some efficiency issues, this is still an open problem, coined the *object identity crisis* by Henry Baker [4]. The interested reader may refer to [15,25].

The transparency, on the other hand, has requirements met by fewer languages. There are two possible approaches to add transparent persistence.

The first one is to use proxy objects, which are objects that mediate any access to the proxied object and serialize and deserialize it according to the type of access. They are only possible when the language provides user-defined implicit conversion operators. C++, for instance, provides them, but with the restriction that in any chain of implicit conversions, only one can be user-defined. For instance, it would be impossible to make transparently persistent objects that are already made transparently versioned with the same method.

Another approach is the use of alternate data structures that exhibit the same interface. This is the approach of a MOP-based persistence library as well as of solutions alike, though those solutions typically restrict themselves to user-defined types. That is, though all types are serializable, transparency is limited to user-defined persistent types. ZODB, a transparent persistence library for Python, takes this approach but, as Python lacks a MOP, uses so-called Extension Classes implemented in C++. To be the least intrusive possible, the class definition syntax should be identical for transient and persistent classes, with the exception of persistence.

### 4.2 Library requirements

Beyond transparent persistence, a MOP-based library needs to fulfill some specific requirements to provide a solution to our problem. In all aspects of CLOS – Common Lisp Object System, persistent objects must exhibit a behaviour consistent with equivalent transient objects. Specifically, two classes, defined by the same `defclass` form but for the persistence metaclass and persistence options, must have the same behaviour, except for MOP-related operations.

Care must be taken that instance creation and initialization and instance deserialization don't interfere, as CLOS explicitly provides generic functions as mechanisms for controlling initialization. If methods have been defined for these generic functions, the persistence library must ensure that they will be called according to standard CLOS semantics.

Class redefinition also constitutes an important issue in itself. As class redefinition needs to operate on the extent of a class, it has an $O(n)$ time complexity on the number of instances of the class. Depending on the size of the data set, if redefinition takes place entirely before any other operation can happen, it could slow the application prohibitively and cause IOS itself, which defeats one of the very purposes of using the library. Concurrent or lazy updates of the instances are possible solutions. In any case, persistence of closures will be necessary for the data to remain consistent, if user-defined methods are defined for the generic functions `update-instance-for-redefined-class` or `update-instance-for-different-class`.

## 5 Using the MOP-based persistence library for dynamicity

Common Lisp includes a number of features that make it particularily suited for rapid prototyping [20]. Prototyping is the gradual building of an application, with the intent to learn from the working implementation. To do rapid prototyping, one needs to be able to build each successive version of the prototype at the least effort, which means being able to make small incremental changes: small incremental additions and small incremental modifications. Those incremental changes are made possible easily and efficiently by the interactive nature of the Lisp image: within the Read-Eval-Print Loop (REPL), without going through a whole edit-compile-link cycle with compilation units as big as entire files, one can:

- add or replace a function or method, possibly compiled on-the-fly,
- execute arbitrary code, also possibly compiled,
- add a new class definition,
- redefine an existing class, which updates all its existing instances.

Persistence fits nicely in this scenario with a MOP-based persistence library: the data model is expanded by adding a new persistent class and modified by redefining an existing persistent class. The developer can experiment interactively with the way data will be stored, create persistent data and test its code against it at a high rate, without any overhead.

The data model no longer needs to be designed entirely before being usable. Parts of it can be gradually built as persistent classes, and any perceived mistake showed by experimentation can be corrected either by defining another class and switching existing instances of the deprecated class to it or by redefining a class, like when it comes to add properties to existing objects, which is easily done by redefining their class with additional slots. If code has not yet been written to use these new slots, the application will even continue to work as before; if code has been written which uses the condition system to deal with unbound slots, the new slots of existing instances can be updated gradually also. This can save the developer a lot of time if he quickly decides to remove some of the new slots, for instance if using them a few times showed their addition was a bad design. This is exactly the kind of benefit in productivity that rapid prototyping can exhibit.

As far as rapid prototyping is concerned, a MOP-based persistence library can also be helpful in a variety of applications regardless of their use of persistent data. Testing an application typically involves using it with a rather complex data set which is more able to show bugs than a simpler one. The problem is, building this data set again and again – especially when it's a complex graph of objects – because it's lost each time the test application is shut down can quickly become tedious. Making this data set persistent, even if it's not meant to be used persistently in production, makes it available in each test run of the application, with little effort.

In any case, when the need to make complex modifications to the data arises, the developer has the full programming language available for the task. Whether he needs to do modifications within an unchanged data model or migrates data when the data model is changed, he suffers no restrictions on how to express the change and on the possible side-effects. This could be some code executed from the REPL directly that would never be used again or factored into a function to be reused later.

The full availability of the language also means that a migration operation is not restricted to data manipulation. Operations like logging or visual rendering of the operation could be added in the code that does the migration. Statistical data could also be extracted.

In production systems as well, Common Lisp shows some rare abilities that come from its dynamicity. In particular, its interactive nature makes it possible to update a running application. This not only minimizes interruption of service (IOS) for software updates, but avoids it fully, provided that the new code is correct. With some infrastructure, rollback and versioning of running code could even be possible. Coupled with automated migration tests [21], this is a great opportunity to avoid a major source of problems in software upgrades.

A MOP-based persistence library makes it possible to update the data model of a running application. As long as the update process is safe and efficient – which can be achieved through lazy updates of existing objects [8], this update can be made without any IOS.

# 6 Experiments and further works

We used the approach described in this paper for the implementation and deployment of a Web-based catalog for real estate. The application was first prototyped, not as a series of independent versions but as a single long-running prototype, incrementally modified to fit the functional requirements of the project. The data model, in particular, evolved through the lifetime of the prototype. In our case, not only did the use and implementation of the prototype give feedback that led the data model to be modified but the client also expressed new or changed requirements.

Without the use of a transparent persistence library, the development of the prototype would not have been possible within the tight time and budget constraints of the project. Moreover, without being concerned by data storage, the work was entirely spent developing and debugging code that implement the logic of the application itself, rather than its supporting infrastructure.

Because of the time constraints of the project, a version of the prototype was asked by the end user to be deployed and used early on. As the development was still taking place, the ability to migrate existing data to the various evolutions of the data model without neither corrupting it nor having to write migration code proved to be absolutely critical.

As a result, once deployed, the system was able to run without any interruption despite the fact that in the meantime its code was heavily modified as was the mere structure of the data it was holding.

The persistence library we use is Elephant, a very high quality MOP-based library with pluggable storage backend. Yet Elephant doesn't already fulfill all requirements described in this paper. In particular, some of these requirements have been devised as we encountered inconsistencies between the expected behaviour of our classes and the one showed by Elephant's persistent classes.

For example, a persistent instance, when deserialized, is allocated by being recreated with the `make-instance` function. As one of our classes had an `:around` method defined for `initialize-instance` with side-effects, those where randomly triggered as deserialization occured, instead of only once, at instance creation. It is worth noting that the needed modifications of the library are in fact relatively easy.

The inconsistency of peristent objects initialization, with respect to transient objects, was encountered as follows. Photographs of products were stored by the application as separate files, along with resized versions, each in numbered files (`1.jpg`, `2.jpg`, etc...). When a photo was added to a product, an object was created with the original photo's file name in a slot. An `:around` method for `initialize-instance` was responsible for the creation of the thumbnail images. Their file names were stored in slots of the object. If the application was a Lisp image running without interruption with transient objects, such file creation would only occur once for each photo. But with persistent objects as they are implemented in Elephant, the `initialize-instance` method was called during object deserialization, and spurious files were created randomly.

Class redefinition also remains a weak point for the library, with respect to our specific requirements. As the library doesn't include the mechanism to ensure that instances currently not deserialized were updated, this had to be triggered manually. This aspect, though, being still an active field of research, will need decisions about the tradeoffs of the implemented solution. Selectable class update mechanisms would provide a very high degree of flexibility and make it possible to better fit the application requirements.

In our specific case, an easy workaround was possible, as all instances were always reachable from in-memory data structures, mostly lists. After class redefinition, all that was needed was to map a function that ensured class migration on the containing lists.

# 7 Conclusion

We have shown how some qualities of Common Lisp, and in particular qualities that together are a specificity of Common Lisp – a dynamic nature enabling hot update of many of an application's components, can be extended to a domain where they initially don't apply. We have seen that such an extension can be made totally transparent for the developer. We have also shown that Common Lisp, along with the MOP, provides such a high degree of flexibility that this extension can be made portably. Although the MOP isn't itself included in the Common Lisp specification, it is a widely provided extension of CLOS.

It is worth noting that a persistence library can find other interesting use cases. In particular, in the case of an application used to edit files in a specified format, the data describing a file might be incorporated in the serialized state to provide faster access to it. It may be stored in its canonical format only when really needed – *e.g.* a working copy or a temporary backup of a text document might be faster to read and parse as part of the serialized state than in the standard format used to communicate documents. Also, when two applications make a consistent use of identical classes and the serialized state is stored in a file, a persistence library could also easily provide an *ad hoc* file format.

## Acknowledgements

We would like to thank the Elephant developing team for their help and support in using and understanding their library.